\documentclass[prl,twocolumn,showpacs]{revtex4}
\usepackage{graphicx}
\begin{document}

\title{
   Magneto-Optical Probing of Weak Disorder 
   in a Two-Dimensional Hole Gas}

\author{
   Leszek Bryja, Arkadiusz W\'ojs, and Jan Misiewicz}
\affiliation{
   \mbox{
   Institute of Physics, Wroclaw University of Technology,
   Wybrze\.ze Wyspia\'nskiego 27, 50-370 Wroclaw, Poland}}

\author{
   Marek Potemski}
\affiliation{
   Grenoble High Magnetic Field Laboratory, CNRS,
   F-38042 Grenoble Cedex 9, France}

\author{
   Dirk Reuter and Andreas Wieck}
\affiliation{
   \mbox{
   Angewandte Festk\"orperphysik, Ruhr-Universit\"at Bochum,
   Universit\"atstrasse 150, 44780 Bochum, Germany}}

\begin{abstract}
In two-beam magneto-photoluminescence spectra of a two-dimensional 
valence hole gas we identify the three-level energy spectrum of a 
free positive trion with a field-induced singlet-triplet transition.
The recombination spectrum of acceptor-bound trions is also detected, 
including a cyclotron replica corresponding to the hole shake-up process.
The emergence of a shake-up peak at low temperature is shown to be 
a sensitive probe of the presence of a small number of impurities
inside the high-mobility quantum well, and its relative position is 
directly related to the hole cyclotron mass.
\end{abstract}
\maketitle

Quantization of single-particle energy into macroscopically 
degenerate, widely separated Landau levels (LLs) makes the 
two-dimensional (2D) gas of charge carriers in a strong 
magnetic field a unique setting for studying the non-perturbative 
many-body interaction phenomena.
Eminent examples of discoveries made in these systems include 
fractional quantum Hall effect \cite{Tsui82}, emergence of 
incompressible fluids with fractional or nonabelian quasiparticles 
\cite{Laughlin83,Halperin84,Moore91}, or formation of topological 
``skyrmion'' excitations \cite{Sondhi93,Barret95} carrying massive 
spin per unit charge.

A powerful tool used to study interacting 2D carriers is 
photoluminescence (PL) spectroscopy \cite{Kukushkin96,Byszewski06}.
In a PL experiment, additional electron--hole ($e$--$h$) pairs 
are introduced into the system through photon absorption, 
and response of the surrounding carriers makes the recombination 
spectrum sensitive to their many-body dynamics.

Since their prediction \cite{Lampert58,Stebe89} and successful 
experimental detection \cite{Kheng93}, trions ($X^\pm=2e+h$ or 
$2h+e$) have been recognized to play a crucial role in the PL 
spectra of 2D gases \cite{Buhmann95,Finkelstein95,Shields95}.
The three-body dynamics of a trion might at first sight seem 
analogous to the familiar Hydrogen ion problem.
However, comparable $e$ and $h$ masses, confinement, magnetic 
field, nonparabolic and anisotropic hole dispersion, and coupling 
to the crystal lattice, surrounding carriers, or impurities -- 
all generate complexity making trions fascinating objects 
of intense experimental \cite{Hayne99,Glasberg99,Yusa01,%
Schuller02,Vanhoucke02,Astakhov05} and theoretical 
\cite{Wojs95,Palacios96,Whittaker97,Dzyubenko00,Wojs00,%
Riva01,Wojs06a,Wojs06b} research.

Nonetheless, the role of trions or other $e$--$h$ complexes 
in magneto-optical spectra is still far from being completely 
understood.
Especially the hole gas remains relatively less thoroughly 
explored because of both more complicated valence band structure 
and a larger difficulty in achieving ultra-high carrier mobility.
The last point invokes the puzzling question of the role played 
by localization or, alternatively, of the influence of weak 
disorder in strongly interacting 2D systems of carriers.

\begin{figure}[t]
\includegraphics[width=3.4in]{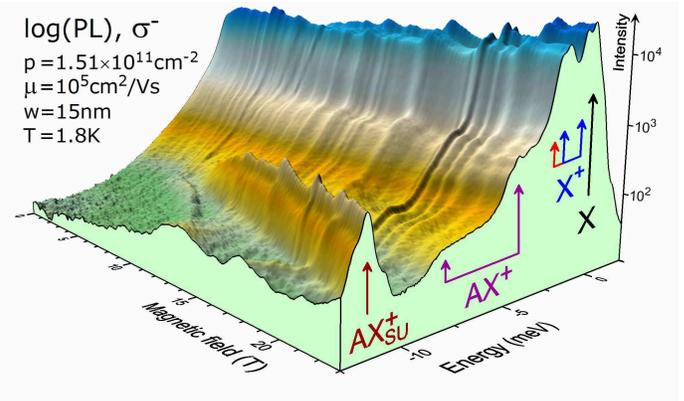}
\caption{(color online)
   The $\sigma^-$ polarized PL spectrum of the 2D valence 
   hole gas as a function of the magnetic field.
   Energy is measured from the exciton line.
   Arrows mark several peaks identified at high fields: 
   exciton $X$, trions $X^+$ (red for spin-singlet and 
   blue for two triplets), and emission spectrum of the 
   acceptor-bound trion $AX^+$, including a shake-up
   line $AX^+_{\rm SU}$.}
\label{fig1}
\end{figure}

This very question is addressed in the present paper.
We report magneto-optical studies of a high-quality 2D hole gas 
in a GaAs quantum well. 
By precise control of experimental conditions we were able to record 
very rich spectra shown in Fig.~\ref{fig1}, including free excitons 
and positive trions, and the transitions involving trions bound to 
residual neutral acceptors inside the well.
In the free trion spectrum, all three anticipated \cite{Wojs00} 
bound states have been identified.
Of those, the positive ``dark triplet'' state $X^+_{\rm td}$ has not, 
to our knowledge, been previously observed.
Crossover to the $X^+_{\rm td}$ ground state has been found at the 
magnetic field $B\approx12$~T.
The variation of energy and intensity of the free trion peaks appears
in coincidence with expected formation of the incompressible hole 
fluids.
The spectrum of impurity-bound trions constitutes a sensitive probe 
of the weak disorder in our sample.
It contains a cyclotron replica corresponding to the ``shake-up'' 
process \cite{Finkelstein96,Glasberg01,Dzyubenko04} in the 
valence band.
Identification of all transitions and especially establishing the 
role of acceptors located inside the well was achieved by 
comparison of experimental data with realistic numerical 
calculations.
Led by prediction that the shake-up splitting is a bare cyclotron 
energy, we were also able to determine the hole cyclotron mass and 
observe an anticrossing of the heavy- and light-hole subbands at 
$B\approx8$~T.

\begin{figure}
\includegraphics[width=3.4in]{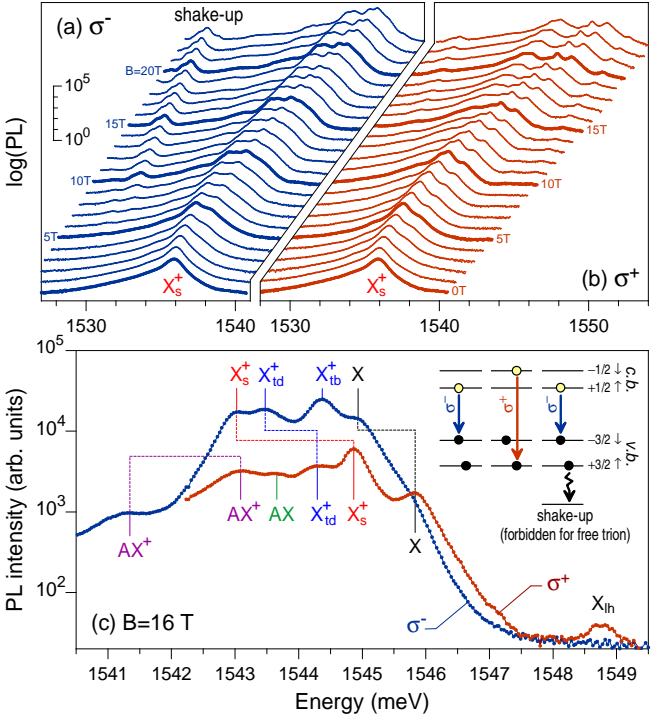}
\caption{(color online)
   Top: Comparison of (a) $\sigma^-$ and (b) $\sigma^+$ polarized 
   PL spectra at a sequence of magnetic fields between $B=0$ and 23~T
   (same sample as in Fig.~\ref{fig1}).
   Bottom: Magnified spectra at $B=16$~T.
   Several identified lines include free exciton $X$ and three positive 
   trions $X^+$ (spin-singlet and two spin-triplets), acceptor-bound 
   exciton $AX$ and trion $AX^+$, and a light-hole exciton $X_{\rm lh}$.
   Note different spin splittings of the $X$, $AX^+$, and different 
   $X^+$ states.
   Inset: Schematic of the $\sigma^\pm$ polarized transitions of 
   a positive spin-singlet trion.}
\label{fig2}
\end{figure}

The studied sample was a GaAs/Al$_{0.35}$Ga$_{0.65}$As quantum 
well of width $w=15$~nm, grown by molecular beam epitaxy on a 
(001) semi-insulating GaAs substrate and symmetrically $\delta$ 
C-doped in the barrier on both sides. 
The concentration and mobility of the holes measured at 
low temperature ($T=4.2$~K) were $p=1.51\cdot10^{11}$~cm$^{-2}$ 
and $\mu=1.01\cdot10^6$~cm$^2$/Vs (the latter corresponding to
the hole mean free path comparable with electron systems at 
$\mu_e\sim10^6$~cm$^2$/Vs).
The PL was excited by a 720~nm red line of Titanium Sapphire 
tunable laser (below the energy gap in the barrier), and an 
additional 514~nm green ion Argon line (exceeding the gap) 
was used to decrease hole concentration. 
The spectra were recorded at $T=1.8$~K and in high magnetic 
fields up to $B=23$~T with a small step $\Delta B=0.1$~T. 
We used Faraday configuration with the linear polarizer and 
wave quater placed together with the sample in liquid helium. 
To switch between the $\sigma^-$ and $\sigma^+$ circular 
polarizations, the magnetic field direction was reversed. 

\begin{figure}
\includegraphics[width=3.4in]{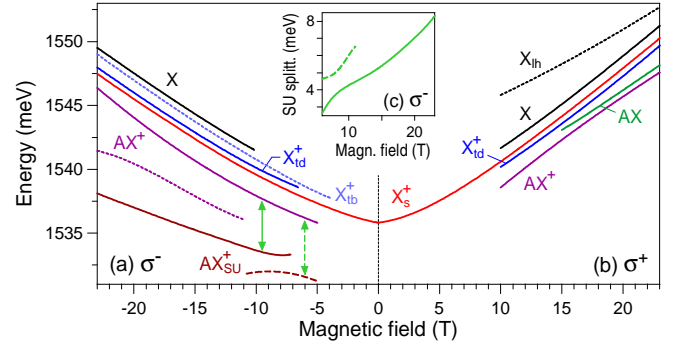}
\caption{(color online)
   Magnetic field dependence of recombination energies 
   identified in polarizations $\sigma^-$ (a) and 
   $\sigma^+$ (b) in the PL spectra in Fig.~\ref{fig2}.
   (c) Splitting of the shake-up line.}
\label{fig3}
\end{figure}

The field evolution of the PL spectrum is presented in 
Fig.~\ref{fig2}(a) and (b).
In the absence of a magnetic field a single line is observed, with 
a characteristic exponential low-energy tail.
It is due to recombination of a spin-singlet trion $X^+_{\rm s}$
(the only bound $2h+e$ state at $B=0$).
More transitions appear when the magnetic field is applied.
Also, a difference emerges between the two polarizations of 
emitted light, corresponding to the recombination of $e$--$h$ pairs 
with different spins ($\uparrow$-$\downarrow$ for $\sigma^-$ and
$\downarrow$-$\uparrow$ for $\sigma^+$).
This is sketched in the inset in Fig.~\ref{fig2}(c), showing the 
spectra for $B=16$~T.
The field dependence of all resolved transition energies is 
displayed in Fig.~\ref{fig3}.

\begin{figure}
\includegraphics[width=3.4in]{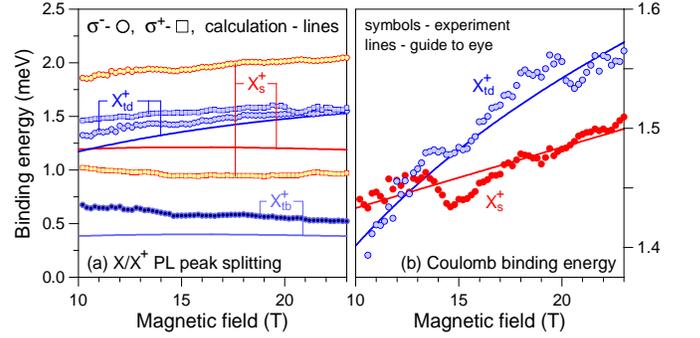}
\caption{(color online)
   (a) Magnetic field dependence of the trion binding 
   energies determined from the exciton--trion peak 
   splittings, $\Delta=E[X]-E[X^+]$;
   lines -- numerical calculation.
   (b) Coulomb binding energies of the singlet and dark 
   triplet trions, obtained as $\sigma^\pm$ averages 
   ${1\over2}(\Delta^++\Delta^-)$ to remove Zeeman effect.
   The crossover occurs at $B\approx12$~T, and adding the 
   hole Zeeman term to $E[X^+_{\rm s}]$ shifts it to an 
   even lower field.}
\label{fig4}
\end{figure}

Among other peaks we identify a neutral exciton $X$ and a pair of 
triplet trions: ``dark'' $X^+_{\rm td}$ and ``bright'' $X^+_{\rm tb}$ 
(the latter only visible in the stronger $\sigma^-$ polarization).
The exciton--trion peak splittings $\Delta$ are plotted as a 
function of $B$ in Fig.~\ref{fig4}(a).
Especially for the $X^+_{\rm s}$, difference between the exciton and 
trion Zeeman splittings must be taken into account to make comparison 
with the numerical Coulomb binding energies drawn with the lines
\cite{Glasberg99,Vanhoucke02,Astakhov05}.
This is done in Fig.~\ref{fig4}(b) by averaging $\Delta$ over both
polarizations, markedly improving agreement with the numerics.
The singlet--triplet crossover (``hidden'' in the PL spectra) is 
revealed at $B\approx12$~T.
The actual ground state transition occurs at a slightly weaker field
due to the hole Zeeman term weakening the $X^+_{\rm s}$ binding.
This prediction matches the observed gradual decrease of the 
$X^+_{\rm s}$ emission beyond $B\approx10$~T.

Convincing identification of the weak lines found on the low-energy 
wing of the trion spectra was achieved by the comparison with 
realistic calculations.
In the high-field $\sigma^+$ spectra, the $AX$ state (an exciton 
bound to a residual neutral acceptor $A=A^-+h$ inside the quantum 
well) is identified.
It is easily distinguished from charged complexes by additional 
application of a green laser which, at sufficient power density, 
converts the structure from $p$- to $n$-type (appropriate green 
illumination also enables detection of the excitonic peak down to 
$B=0$).
The pair of $\sigma^\pm$ lines labeled as $AX^+$ are attributed to 
the recombination of a trion bound to a neutral acceptor.
At an even lower energy, a cyclotron replica of the $AX^+$ peak was
detected.
It describes a shake-up process \cite{Finkelstein96,Glasberg01} 
in which the $e$--$h$ annihilation is accompanied by excitation 
of a left-over acceptor-bound hole to a higher LL.

Remarkably, the selection rules associated with translational symmetry 
of the quantum well preclude shake-up recombination of free trions
\cite{Dzyubenko04}. 
Furthermore, we found that shake-up which arises from scattering 
of trions by free carriers is virtually negligible at fractional LL 
fillings due to Laughlin trion--carrier correlations \cite{Wojs06a}.
Also in the comparable electron systems shake-up is much weaker due 
to a larger cyclotron gap.
These facts make our system well suitable for studying the shake-up 
effect.
More importantly, its detection is a sensitive probe of a small number 
of residual impurities in a high-quality structure.

Calculation consisted of exact diagonalization of model hamiltonians 
of small $e$--$h$ systems, with and without an additional point charge 
$A^-$ in the middle of the quantum well (other near-center positions 
gave equivalent results, but placing $A^-$ closer to the well's edge 
caused qualitatively different behavior which could not explain the 
experimental spectra).
Two spin states and up to five LLs and two subbands were included for 
electrons and holes.
Coulomb matrix elements were integrated exactly, using realistic 
3D subband wave functions.
Further details were recently given in Ref.~\cite{Wojs06b} and will 
not be repeated here.

\begin{figure}
\includegraphics[width=3.4in]{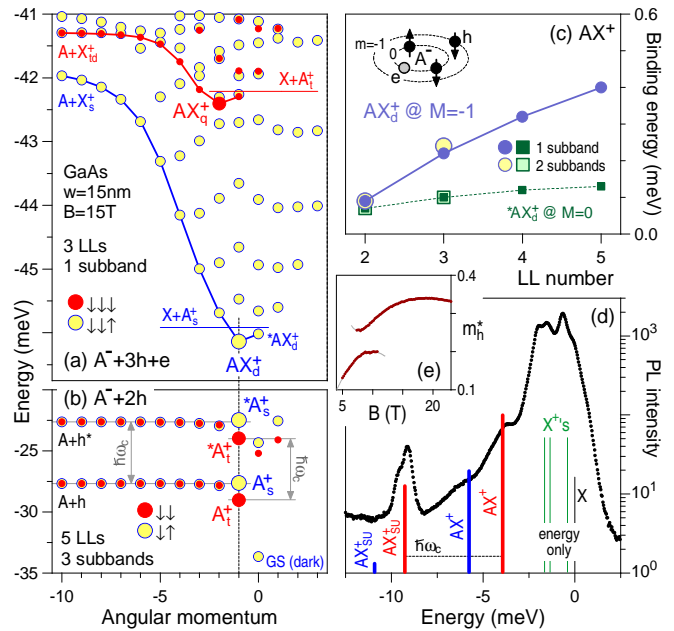}
\caption{(color online)
   Calculated energy spectra of (a) $3h+e$ and (b) $2h$ systems 
   in the presence of an ionized acceptor $A^-$ in the middle of 
   a symmetric quantum well.
   In (a), $AX^+_{\rm d}$ at angular momentum $M=-1$ is the 
   spin-doublet ground state of an $A$-bound trion.
   In (c), its binding energy is compared to the weakly bound 
   $^*\protect\rule{-1mm}{0mm}AX^+_{\rm d}$ at $M=0$ as a function 
   of the number of included LLs and subbands.
   In (b), four lowest final states for the $AX^+_{\rm d}\rightarrow 
   A^+$ recombination are marked at $M=-1$.
   In (d), the corresponding four-peak $AX^+_{\rm d}$ emission
   spectrum is overlaid with the experimental data of 
   Fig.~\ref{fig2}(a).
   In (e), the hole effective mass $m_h^*$ is extracted from 
   the splitting of the experimental shake-up line, equal to 
   the hole cyclotron energy as predicted in (b).
   All data are for a 15~nm GaAs quantum well and, 
   except for (e), for magnetic field $B=15$~T.}
\label{fig5}
\end{figure}

Let us summarize the numerical results.
First, the Coulomb binding energy of 2.65~meV was obtained (including 
five LLs and two well subbands) for the neutral $AX=A^-+2h+e$ state 
at $B=15$~T.
Assumming the $AX$ Zeeman splitting similar to $X^+$ and $AX^+$, 
this matches perfectly the observed relative position of $AX$ 
in the PL spectrum.
In the next step, a more strongly bound charged $AX^+=A^-+3h+e$ state 
was found in the presence of excess holes.
It was established as the most stable radiative complex in the presence 
of an $A^-$ (with the $AX$ prevailing for the $A^-$ located outside the 
well).
An example of the $A^-+3h+e$ spectrum is presented in Fig.~\ref{fig5}(a).
The ground state is a spin-unpolarized (doublet) state $AX^+_{\rm d}$
with angular momentum $M=-1$ corresponding to the ``compact'' 
single-particle configuration shown in the inset in Fig.~\ref{fig5}(c).
The $AX^+_{\rm d}$ has much lower energy than a trion unbound 
from the $A$ (the $A$--$X^+$ interaction pseudopotential is drawn 
with a solid line).
However, it is rather weakly bound against the break-up into a 
spin-singlet $A^+_{\rm s}$ and a free exciton.
Another marginally bound state $^*\rule{-1mm}{0mm}AX^+_{\rm d}$ 
occurs at $M=0$.
All spin-polarized (quartet) states have higher energy, even
with the Zeeman term, and thus can be neglected.

Fig.~\ref{fig5}(c) compares the $AX^+_{\rm d}$ and 
$^*\rule{-1mm}{0mm}AX^+_{\rm d}$ binding energies $\Delta$ computed 
more accurately by including more LLs or well subbands.
Clearly, only the $AX^+_{\rm d}$ is expected to show in PL, and its 
small $\Delta\sim0.5$~meV makes the detection depend on low temperature
(indeed, only the free $X$ and $X^+$ are observed in our experiment
above $T\sim5$~K).

The accurate energy spectrum of $A^+$ (i.e., of the final state in 
the $AX^+_{\rm d}$ recombination) is shown in Fig.~\ref{fig5}(b).
The four lowest $A^+$ states in the optically active $M=-1$ subspace
(defined by the $\Delta M=0$ selection rule) include two singlets 
and two doublets, derived from single-particle configurations with 
either both holes in the lowest LL ($A^+$) or with one hole in the
higher LL ($^*\rule{-1mm}{0mm}A^+$).

The oscillator strengths for the $AX^+_{\rm d}\rightarrow 
A^+_{\rm s}$, $A^+_{\rm t}$, $^*\rule{-1mm}{0mm}A^+_{\rm s}$, 
and $^*\rule{-1mm}{0mm}A^+_{\rm t}$ transitions were calculated 
to predict the $AX^+$ recombination spectrum in Fig.~\ref{fig5}(d).
Coulomb transition energies obtained from exact diagonalization 
have been additionally shifted by the Zeeman correction determined 
from Fig.~\ref{fig2}(c).
Good agreement with the experimental PL spectrum supports 
identification of the $AX^+$ lines.

Remarkably, the energy distance between $A^+_{\rm t}$ and 
$^*\rule{-1mm}{0mm}A^+_{\rm t}$ appears nearly identical to 
the hole cyclotron gap $\hbar\omega_c$ over a wide range of
magnetic fields.
This means equal binding between the neutral acceptor and 
a hole in the lowest or excited LL.
More importantly, it allows extraction of the hole cyclotron 
mass $m_h^*$ from the position of a shake-up line in the PL 
spectrum.
The result for our experimental data shown in Fig.~\ref{fig5}(e)
reveals an anticrossing of the heavy- and light-hole subbands 
at $B\approx8$~T.

In conclusion, using polarization-resolved magneto-PL of 
a 2D hole gas we detected all three states of a free positive 
trion (singlet, dark triplet, and bright triplet), earlier 
reported only for negative trions.
The ``hidden'' singlet--triplet crossover is found at a 
rather weak field $B\approx12$~T.
We also observed correlation between the changes in energy and 
intensity of the trion recombination and condensation of the 
holes into a series of incompressible fluids, thus confirming 
high quality of the studied hole gas.
Nevertheless, additional lower-energy PL lines could not be 
explained by assuming perfect translational symmetry.
By combining experiment with analysis of optical selection 
rules and realistic numerics, we attributed these lines to the 
acceptor-bound trions, thus establishing PL
as a sensitive probe of the weak disorder in our sample.
A shake-up peak was also identified in the spectra, with the
relative position unaffected by interactions and thereby 
directly linked to the hole cyclotron mass.

Work supported by research grants: N20210431/0771 of the Polish 
MNiSW and RITA-CT-2003-505474 of EC.

\end{document}